\documentclass[twocolumn,traditabstract]{aa}
\usepackage{txfonts}
\usepackage{graphicx}
\usepackage{natbib}
\newcommand{\kms}{\mbox{km~s$^{-1}$}}
\newcommand{\mols}{\mbox{molec.~s$^{-1}$}}
\newcommand{\htwoo}{H$_2^{16}$O~}
\newcommand{\htwood}{H$_2^{18}$O~}
\newcommand{\nratio}{$^{14}$N/$^{15}$N~}
\newcommand{\sratio}{$^{32}$S/$^{34}$S~}
\newcommand{\cratio}{$^{12}$C/$^{13}$C~}
\newcommand{\oratio}{$^{16}$O/$^{18}$O~}

\begin{document}

\title{Isotopic ratios of H, C, N, O, and S in comets C/2012~F6 (Lemmon) and 
C/2014 Q2~(Lovejoy)  
\thanks{Based on observations carried out with the IRAM 30m telescope.
IRAM is supported by INSU/CNRS (France), MPG (Germany) and IGN (Spain).}
\thanks{Odin is a Swedish-led satellite project funded jointly 
        by the Swedish National Space Board (SNSB), the Canadian Space Agency 
        (CSA), the National Technology Agency of Finland (Tekes) and the
        Centre National d'\'Etudes Spatiales (CNES, France). 
        The Swedish Space Corporation is the
        prime contractor, also responsible for Odin operations.}
\thanks{The spectra datset is available at the CDS via anonymous 
ftp to cdsarc.u-strasbg.fr (130.79.128.5)
or via http://cdsweb.u-strasbg.fr/cgi-bin/qcat?J/A+A/}}

\author{N. Biver\inst{1}
   \and R. Moreno\inst{1}
   \and D. Bockel\'ee-Morvan\inst{1}
   \and Aa. Sandqvist\inst{2}
   \and P. Colom\inst{1}
   \and J. Crovisier\inst{1}
   \and D.C. Lis\inst{3}
   \and J. Boissier\inst{4}
   \and V. Debout\inst{1}
   \and G. Paubert\inst{5}
   \and S. Milam\inst{6}
   \and A. Hjalmarson\inst{7}
   \and S. Lundin \inst{8} 
   \and T. Karlsson\inst{8} 
   \and M. Battelino\inst{8} 
   \and U. Frisk\inst{9}
   \and D. Murtagh\inst{10} 
   \and the Odin team.}

\institute{LESIA, Observatoire de Paris, PSL Research University, CNRS, 
Sorbonne Universit\'es, UPMC Univ. Paris 06, Univ. Paris Diderot, 
Sorbonne Paris Cit\'e, 5 place Jules Janssen, F-92195 Meudon, France
   \and Stockholm Observatory, AlbaNova University Center, 
        SE-106 91 Stockholm, Sweden
   \and LERMA, Observatoire de Paris, PSL Research University, CNRS, 
        Sorbonne Universit\'es, UPMC Univ. Paris 06, 
        F-75014, Paris, France
   \and IRAM, 300, rue de la Piscine, F-38406 Saint Martin d'H\`eres, France
   \and IRAM, Avd. Divina Pastora, 7, 18012 Granada, Spain
   \and NASA Goddard Space Flight Center, Astrochemistry Laboratory, Code 691.0, Greenbelt, MD 20771, USA
   \and Onsala Space Observatory, Chalmers University of Technology, SE-439 92 ONSALA, Sweden
   \and OHB Sweden, P.O. Box 1269, SE-164 29 Kista, Sweden
   \and Omnisys Instruments, August Barks Gata 6B, SE-421 32 V\"{a}stra Fr\"{o}lunda, Sweden
   \and Dept. of Radio and Space Science, Chalmers Technical University, Gothenburg, Sweden
}

   \titlerunning{Isotopic ratios of H, C, N, O, S in comets Lemmon and Lovejoy}
   \authorrunning{Biver et al.}
   \date{\today}

   \abstract{The apparition of bright comets C/2012~F6 (Lemmon) and 
C/2014~Q2 (Lovejoy) in March-April 2013 and January 2015, combined with 
the improved observational capabilities of submillimeter facilities, 
offered an opportunity to carry out sensitive compositional and isotopic 
studies of the volatiles in their coma. 
We observed comet Lovejoy with the IRAM 30m telescope between 13 and 26 
January 2015, and with the Odin submillimeter space observatory on 
29 January - 3 February 2015. We detected 22 molecules and several 
isotopologues. The \htwoo and \htwood production rates measured with Odin 
follow a periodic pattern with a period of 0.94 days and an amplitude 
of $\sim$25\%.  
The inferred isotope ratios in comet Lovejoy are $^{16}$O/$^{18}$O=$499\pm24$ 
and D/H =$1.4\pm0.4\times10^{-4}$ in water, 
$^{32}$S/$^{34}$S = $24.7\pm3.5$ in CS, 
all compatible with terrestrial values. 
The ratio $^{12}$C/$^{13}$C = $109\pm14$ in HCN is 
marginally higher than terrestrial and $^{14}$N/$^{15}$N = $145\pm12$ 
in HCN is half the Earth ratio. 
Several upper limits for D/H or \cratio in other molecules
are reported. From our observation of HDO in comet C/2014~Q2 (Lovejoy), 
we report the first D/H ratio in an Oort Cloud comet 
 that is not larger than the terrestrial value. 
On the other hand, the observation of the same HDO line in the other Oort-cloud 
comet, C/2012~F6 (Lemmon), suggests a D/H value 
four times higher. Given the previous measurements of D/H in cometary water,
this illustrates that a diversity in the D/H ratio and in the chemical composition, 
is present even within the same dynamical 
group of comets, suggesting that current dynamical groups contain comets 
formed at very different places or times in the early solar system.}


\maketitle

\section{Introduction}
Comets are the most pristine remnants of the formation of the
solar system 4.6 billion years ago. Investigating the
composition of cometary ices provides clues to the physical
conditions and chemical processes at play in the primitive solar
nebula. Comets may also have played a role in the delivery of
water and organic material to the early Earth 
\citep[see][ and references therein]{Har11}.
The latest simulations of the early solar system evolution \citep{Bra13,Obr14},
suggest a more complex scenario. On the one hand, ice-rich bodies formed beyond
Jupiter may have been implanted in the outer asteroid belt and participated in
the supply of water to the Earth or, on the other hand, current comets coming
from either the Oort Cloud or the scattered disk of the Kuiper belt may have
formed in the same trans-Neptunian region sampling the same diversity of 
formation conditions. Understanding the diversity in composition and isotopic
ratios of the comet material is thus essential in order to assess such
scenarios \citep{Alt03,Boc15}.  

The recent years have seen significant improvement in the sensitivity
and spectral coverage of millimeter receivers. The EMIR
receivers \citep{Car12} at the Institut de radioastronomie millim\'etrique
(IRAM) are equipped with a fast Fourier transform
spectrometer that offers a wide frequency coverage at a high
spectral resolution (0.2~MHz). The combination enables sensitive
spectral surveys of cometary atmospheres and simultaneous 
observations of several molecules, including isotopologues
in brighter comets. We report here observations of isotopic ratios with 
the IRAM~30m radio telescope in two very active (maximum water outgassing 
rate close to $10^{30}$~\mols) Oort cloud comets, C/2012~F6 (Lemmon) and 
C/2014~Q2 (Lovejoy), carried out in 2013 and 2015. They are 
both dynamically old Oort cloud comets: an original orbital period of 
9\,800 years and an orbit inclination of 83\degr~ for comet Lemmon, 
and a period of 11\,000 years and an inclination of 80\degr~ for comet Lovejoy.

The analysis of the observation in terms of molecular abundances and 
detection of rare and new molecular species has been reported 
\citep{Biv14,Biv15}. In the present article we concentrate on the
measurement of several isotopic ratios.
We report the detection of HDO in both comets. The compounds \htwoo and \htwood 
were detected in comet Lovejoy
with the Odin submillimeter space observatory, which helped to accurately determine 
 the D/H and \oratio ratio in water in this comet.

\section{Observations}
\subsection{IRAM Observations}

Comet C/2014~Q2 (Lovejoy) was observed with the IRAM 30m radio telescope
during two periods: on 13.8, 15.8, and 16.8 January 2015 (geocentric distance 
$\Delta=0.496-0.528$ AU) and on 23.7, 24.7, 25.7, and 26.7 January 2015 
($\Delta=0.624-0.675$ AU) under very good weather conditions. 
The heliocentric distances were 1.31 and 1.29 AU, respectively. Perihelion was 
on 30.07 January UT, at 1.290 AU from the Sun. 
The 13--25 and 26 January observations were conducted with
the EMIR 1mm and 3mm receivers, respectively. The main backend used 
is a Fourier transform spectrometer, which covers 
a frequency range of 2$\times$8~GHz (two sidebands separated by 8~GHz,
each in two linear polarizations) in a single setup, with a high spectral 
resolution of 200~kHz. The spectral resolution (0.3--0.2~\kms when converted 
into Doppler velocity) allows the velocity profiles of the narrow 
cometary lines ($\sim2$~\kms) to be resolved.
We also used the high resolution VESPA autocorrelator set to 40~kHz sampling 
on dedicated lines of interest (e.g., CH$_3$OH~($5_{+2}-4_{+1}E$)
and H$^{15}$CN~(3-2)), but VESPA units can only cover lines within a 0.5~GHz
window centered at 6.25~GHz in the IF band (upper and lower side band).
Using four different tunings we covered the 210--218, 225--233, and 
240--272~GHz frequency ranges and 85--93 and 101--109~GHz.

The half power beam width of the IRAM 30~m telescope in the 1~mm band ranges from 
9.1\arcsec~ to 11.6\arcsec, which corresponds to 3300 to 5400~km at 
the comet distance
(22--28\arcsec~ at 3~mm corresponding to 11000--14000~km on 26.7 Jan.). 
Given the expansion velocity of the coma ($\sim0.8$~\kms) and 
heliocentric distance, this means that we are not very sensitive to the
photo-dissociation rate of the molecules, provided that the molecule lifetime
at 1 AU is larger than 4000~s. The pointing accuracy (1\arcsec-2\arcsec) 
was regularly checked on reference pointing sources and also 
on the comet with coarse maps.
The time variation of the activity of the comet could be monitored with the
multiple strong CH$_3$OH lines present in all setups. We did not see
any variations significantly larger than $\pm20$\% 
during the 
observations ($Q_{CH_3OH}=1.2\pm0.2\times10^{28}$~\mols, 13--16 January, 
and $1.4\pm0.2\times10^{28}$~\mols, 23--25 January), which justified the averaging of several days of observations.

Comet C/2012~F6 (Lemmon) was observed in March and April 2013 
\citep[as already presented in][]{Biv14}. Perihelion was on 24.51 March
2013 UT at 0.731 AU from the Sun. 
We mostly used the same tunings as for comet Lovejoy with the EMIR 1mm
receiver combined with the FTS and VESPA spectrometers. The best weather
conditions were found on 14.5 and 18.5 March, but with a comet at low
elevation (15-20\degr), and on 6.5 April 2013. We covered the 240--272~GHz 
frequency range and obtained noisier data on
the 85--93, 166--170, and 210--242~GHz bands with the EMIR 3mm, 2mm, and 1mm
receivers, respectively.
A summary of the observational setups is given in Table~\ref{tabsetup}.

In this paper we focus only on the measurement of isotopic ratios, and 
spectra of isotopologues are presented in Figs.~\ref{figh2oisotopesq2}--
\ref{figcsisotopesf6}. The chemical inventory of these comets based on these
observations with the IRAM telescope has been presented in \citet{Biv14,Biv15}
and will be further detailed in a future paper.

\begin{table*}
\caption[]{Circumstances of IRAM observations and reference parameters.}\label{tabsetup}\vspace{-0.2cm} \centering
\begin{tabular}{lcccccccc}
\hline\hline\noalign{\smallskip}
UT date  & $<r_{h}>$  & $<\Delta>$   & Integ. time & Freq. range & $v_{exp}$  & $T_{kin}$ & $Q_{\rm H_2O}^{b,c}$ & $Q_{\rm CH_3OH}$ \\
$($yyyy/mm/dd.d--dd.d) & (AU)  & (AU)  & (min)$^a$ & (GHz) & (\kms)  & (K) & (\mols) & (\mols) \\
\hline\noalign{\smallskip}
\multicolumn{8}{l}{Comet C/2012~F6 (Lemmon):} \\
2013/03/14.52--14.55 & 0.758 & 1.348 &  26 & 249-256, 264-272 & 1.10 & 110 & $\sim10\times10^{29}$ & $2.1\pm0.1\times10^{28}$ \\
2013/03/14.57--14.59 & 0.758 & 1.349 &  20 & 240-248, 256-264 & 1.10 & 110 & $\sim10\times10^{29}$ & $2.3\pm0.1\times10^{28}$ \\
2013/03/15.52--15.57 & 0.753 & 1.362 &  21 & 249-256, 264-272 & 1.10 & 110 & $\sim10\times10^{29}$ & $1.7\pm0.2\times10^{28}$ \\
2013/03/15.58--15.59 & 0.753 & 1.362 &  14 & 210-218, 226-234 & 1.10 & 110 & $\sim10\times10^{29}$ &                         \\
2013/03/18.53--18.55 & 0.741 & 1.400 &  17 & 249-256, 264-272 & 1.10 & 110 & $\sim10\times10^{29}$ & $1.6\pm0.2\times10^{28}$ \\
2013/03/18.55--18.59 & 0.741 & 1.400 &  20 & 240-248, 256-264 & 1.10 & 110 & $\sim10\times10^{29}$ & $2.1\pm0.1\times10^{28}$ \\
2013/04/06.40--06.51 & 0.776 & 1.592 &  69 & 249-256, 264-272 & 1.00 & 100 & $\sim9\times10^{29}$ & $1.5\pm0.1\times10^{28}$ \\
2013/04/06.53--06.60 & 0.776 & 1.592 &  54 & 240-248, 256-264 & 1.00 & 100 & $\sim9\times10^{29}$ & $1.4\pm0.1\times10^{28}$ \\
2013/04/07.49--07.52 & 0.783 & 1.600 &  20 & 249-256, 264-272 & 1.00 & 100 & $\sim9\times10^{29}$ & $1.6\pm0.2\times10^{28}$ \\
2013/04/07.53--07.59 & 0.783 & 1.600 &  58 & 210-218, 226-234 & 1.00 & 100 & $\sim9\times10^{29}$ & $1.2\pm0.4\times10^{28}$ \\
2013/04/08.43--08.50 & 0.790 & 1.606 &  62 &  85-93,  166-170 & 1.00 & 100 & $\sim9\times10^{29}$ & $1.5\pm0.1\times10^{28}$ \\
2013/04/08.51--08.54 & 0.790 & 1.606 &  28 & 249-256, 264-272 & 1.00 & 100 & $\sim9\times10^{29}$ & $1.5\pm0.1\times10^{28}$ \\
2013/04/08.55--08.59 & 0.791 & 1.607 &  36 & 218-226, 234-242 & 1.00 & 100 & $\sim9\times10^{29}$ & $1.7\pm0.2\times10^{28}$ \\
\multicolumn{8}{l}{Comet C/2014~Q2 (Lovejoy):} \\
2015/01/13.75--13.77 & 1.314 & 0.496 &  14 & 249-256, 264-272 & 0.80 & 73 & $5\times10^{29}$ &  $1.17\pm0.03\times10^{28}$ \\
2015/01/13.79--13.84 & 1.314 & 0.497 &  29 & 210-218, 225-233 & 0.80 & 73 & $5\times10^{29}$ &  $1.25\pm0.07\times10^{28}$ \\
2015/01/13.85--13.88 & 1.313 & 0.497 &  29 & 240-248, 256-264 & 0.80 & 73 & $5\times10^{29}$ &  $1.25\pm0.02\times10^{28}$ \\
2015/01/15.82--15.83 & 1.308 & 0.516 &  13 & 249-256, 264-272 & 0.80 & 73 & $5\times10^{29}$ &  $1.18\pm0.04\times10^{28}$ \\
2015/01/15.85--15.96 & 1.308 & 0.516 & 105 & 240-248, 256-264 & 0.80 & 73 & $5\times10^{29}$ &  $1.42\pm0.02\times10^{28}$ \\
2015/01/16.76--16.78 & 1.306 & 0.526 &  14 & 249-256, 264-272 & 0.80 & 73 & $5\times10^{29}$ &  $1.14\pm0.01\times10^{28}$ \\
2015/01/16.81--16.95 & 1.306 & 0.527 & 120 & 240-248, 256-264 & 0.80 & 73 & $5\times10^{29}$ &  $1.17\pm0.02\times10^{28}$ \\
2015/01/23.72--23.75 & 1.294 & 0.624 &  36 & 240-248, 256-264 & 0.80 & 73 & $6\times10^{29}$ &  $1.48\pm0.03\times10^{28}$ \\
2015/01/24.73--24.75 & 1.293 & 0.641 &  14 & 226-234, 241-249 & 0.80 & 73 & $6\times10^{29}$ &  $1.27\pm0.06\times10^{28}$ \\
2015/01/25.71--25.75 & 1.292 & 0.658 &  42 & 210-218, 225-233 & 0.80 & 73 & $6\times10^{29}$ &  $1.00\pm0.04\times10^{28}$ \\
2015/01/26.74--26.75 & 1.291 & 0.675 &   6 &  85-93, 101-109 & 0.80 & 73 & $7\times10^{29}$ &  $1.99\pm0.20\times10^{28}$ \\
\hline
\end{tabular}\vspace{-0.2cm}
\tablefoot{$^a$ On nucleus pointings only.\\
$^{b}$ For comet Lemmon, \citet{Com14} provide $Q_{\rm H_2O}$ values in the range
7--12$\times10^{29}$\mols during this period. Contemporaneous SWAN measurements
are slightly lower than quoted here, but they were not time-deconvolved and the 
average values measured by SWAN a few days later are within 10\% of those given here. 
Given our measured $Q_{\rm CH_3OH}$ (Col. 9), $Q_{\rm H_2O}$ are also in agreement with the 
measurement of \citet{Pag14} as they yield the same $Q_{\rm CH_3OH}$/$Q_{\rm H_2O}$ ratio.
Nan\c{c}ay observations are also compatible with these $Q_{\rm H_2O}$ 
although the quenching of the OH maser is poorly constrained and the uncertainty 
on $Q_{\rm OH}$ is very large.\\
$^c$ For comet Lovejoy, $Q_{\rm H_2O}$ are based on Odin and Nan\c{c}ay data 
(see text and \citet{Biv15}).}
\end{table*}

\subsection{Odin observations of comet C/2014~Q2 (Lovejoy)}

Because comet C/2014~Q2 (Lovejoy) clearly became a very active comet, 
observations with the Odin 1.1m submillimeter satellite \citep{Fri03} 
were triggered on a late notice. Observations took place between 30.5 January 
and 3.4 February 2015. These observations were scheduled to map the 
emission of \htwoo at 556.9~GHz, follow its temporal evolution, and 
detect the \htwood line at 547.7~GHz. The half power beamwidth is close 
to 2.1\arcmin~ at these frequencies and the main beam efficiency is 0.89.
Odin uses three spectrometers:
a 1 GHz bandwidth acousto-optical spectrometer (AOS with 0.6~MHz sampling)
and two autocorrelators (AC1 and AC2) used in their highest resolution mode 
($\sim$0.15~MHz with 100~MHz bandwidth). For the first setup, two receivers 
tuned to the \htwoo line were used in parallel, coupled to the three 
spectrometers: AOS and AC2 on receiver 555-B2 and AC1 on 549-A1.
For the second setup, the AOS was connected to the 572-B1
receiver tuned near the ammonia line at 572.5~GHz, but not frequency locked,
while AC2 was connected to the receiver 555-B2 and AC1 to 549-A1, 
both tuned to the \htwood line. 

As Odin is in a helio-synchronous polar Earth orbit at 550~km altitude, 
the comet was only observable  during $\sim$55~min of each 96~min orbit.
The Earth was in the field of view during the remaining time, and
the atmospheric lines were used to calibrate the frequency scale.
Some of the observations failed to track  the comet correctly, 
but half were successful and provided good results.
Spectra of the nucleus-centered \htwoo and \htwood
lines are shown in Fig.~\ref{figh2oisotopesq2} and a map of the \htwoo 
line is shown in Fig.~\ref{maph2o}.

\section{Modeling of the lines and model parameters}
\subsection{Velocity and outgassing pattern}

The mean gas expansion velocity was derived from the shape of the lines 
with the highest signal-to-noise ratio. 
We determined a value of 1.1 to 1.0~\kms~ for 
comet Lemmon and 0.8~\kms for comet Lovejoy (Table~\ref{tabsetup}). 
As discussed hereafter,  comet Lovejoy was observed with 
the larger beam of Odin; a value of 0.85~\kms~ provided a better fit to the 
width of the water lines. 
This is expected, as acceleration of the gas expansion velocity 
in the coma is predicted and observed in the outer part of the
comae of active comets. 

\subsection{Temperature and collisional excitation}

For all molecules we used the latest spectroscopic data available
in the JPL \citep{JPLM} and CDMS \citep{CDMS} databases, both for line
identification and computation of line strengths. 
We employed the code previously used for other 
comets \citep{Biv07,Har11,Boc12} to compute the 
excitation of the rotational levels of the molecules and the line intensities. 
We take into account the pumping of the rotational levels by the 
fluorescence of the infrared vibrational bands, which 
is not expected to play a major role at the cometocentric distances
$\leq$5000~km sampled in the case of comet Lovejoy. 
In our model molecules slowly evolve from the local thermal 
equilibrium ($T=100-110$ and 73 K, for comets Lemmon and Lovejoy, respectively; Table~\ref{tabsetup})
maintained by collisions with neutrals and electrons close to the nucleus 
to spontaneous decay of the rotational population in the 
less dense parts of the coma \citep{Biv00,Biv06}. Within the accuracy of the detections,
line intensity ratios (or rotational temperatures) for methanol, and
also for most other species, are in agreement
with the model (e.g., $T_{rot}$(CH$_3$CHO, Lovejoy)$=67\pm15$ K for 68 K 
predicted). The density is described with the Haser model, using lifetimes from 
\citet{Cro94,Boc00,Cro04a}; and \citet{Cro04b}. 
Line intensities are computed with a 
radiative transfer code that takes  opacity
effects into account. Except for \htwoo and HCN, lines are optically thin, meaning that 
line intensities are proportional to the production rates.

\section{Water production rate and time evolution}

\begin{figure*}[ht]
\centering
\resizebox{\hsize}{!}{\includegraphics[angle=270]{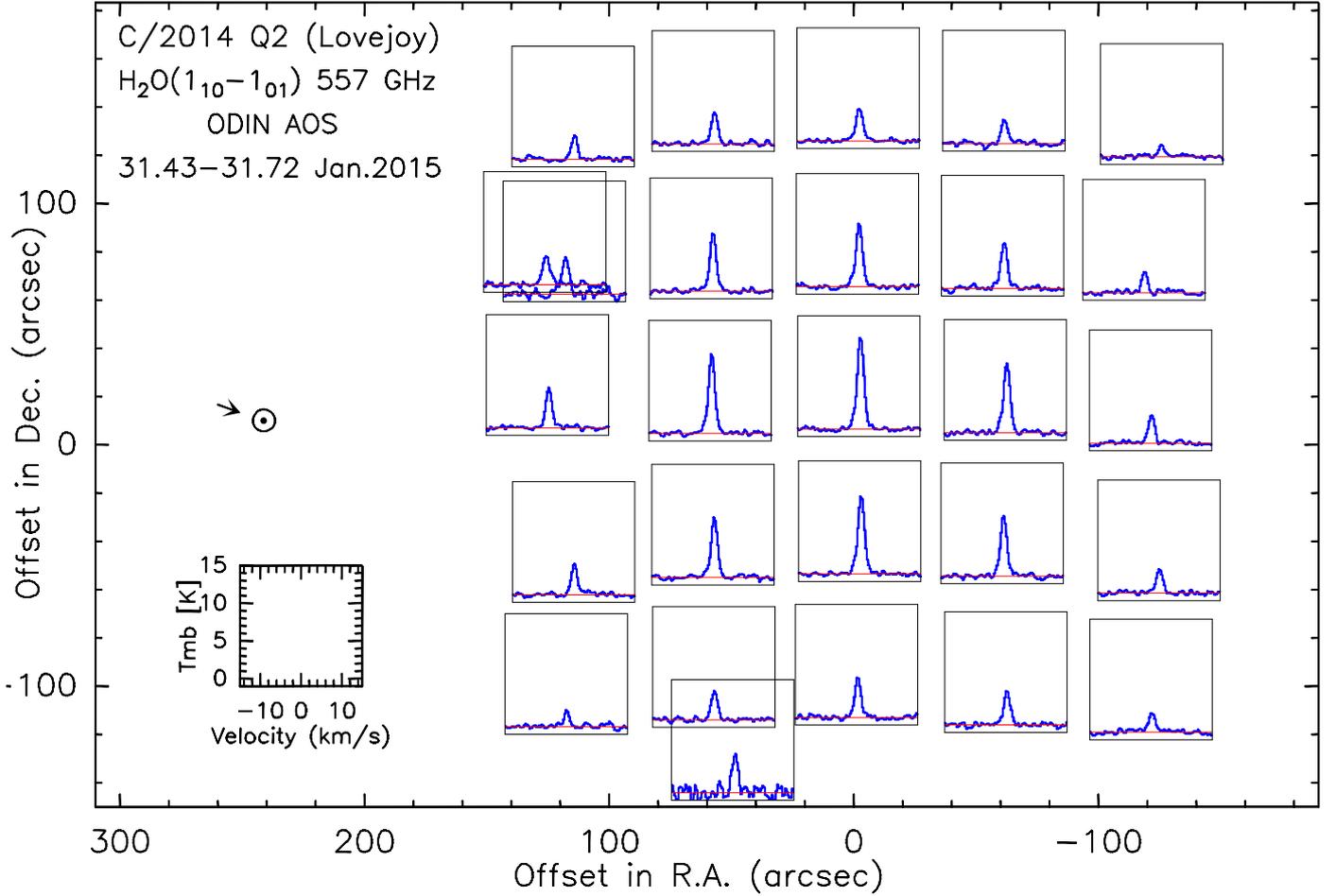}}
\caption{Map of the \htwoo~($1_{10}-1_{01}$) line obtained with Odin {\bf and 
its acousto-optical spectrometer} in comet Lovejoy. The arrow provides 
the projected direction of the Sun (phase angle was 50\degr), but the water 
emission does not exhibit any significant asymmetry.}
\label{maph2o}
\end{figure*}

\subsection{Odin maps and excitation of water in the coma of comet Lovejoy}

Several maps (e.g., Fig.~\ref{maph2o}) of the water emission were performed 
for comet Lovejoy with Odin in order to precisely derive the water
production rate.
The extension of the emission of optically thick submillimeter water lines
in cometary comae is in large part governed by the excitation mechanism
of the water rotational levels. It  depends both on the temperature profile
and collision rates with electrons and neutrals in the inner part of the coma
\citep{Biv07,Har11,Boc12}. However, in the outer part (beyond 150\arcsec) the
line intensity mostly depends on the water production rate (and assumed 
photo-dissociation lifetime), and very little on the assumed temperature 
profile. Dividing the gas temperature by a factor of 2 only increases
the derived production rate  by
10\%. We used the same excitation parameters
for the \htwoo observations with Odin, as those derived for the molecular 
lines observed with the IRAM 30m instrument: a constant temperature of 73 K throughout 
the coma and an electron density scaling factor $x_{ne} = 0.5$ 
\citep{Zak07,Biv07}. The expansion velocity inferred from the 
\htwoo~ and \htwood~ line widths is $v_{exp}=0.85$~\kms~, slightly larger
than for the lines observed with the smaller beam of the IRAM 30m telescope.
The fit to the radial profile of the line intensity from the map 
(Fig.~\ref{maph2o}) is satisfactory and corresponds to
$Q_{\rm H_2O}=8.7\pm0.7\times10^{29}$~\mols
and the reduced $\chi^2_{\nu}=2.4$ when we include 
the 5\% uncertainty due to the 10\arcsec ~\citep{Fri03} pointing uncertainty. 
Taking into account an increase in the velocity with distance from the nucleus
does not significantly change the goodness of the fit and the derived 
production rate.

\begin{figure}
\centering
\resizebox{\hsize}{!}{\includegraphics[angle=270]{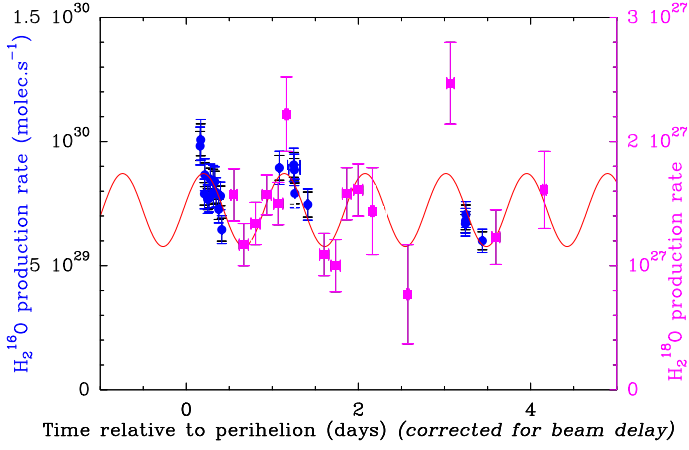}}
\caption{\htwoo (left vertical scale) and \htwood (pink, right vertical scale) 
production rates in comet Lovejoy from Odin observations. Horizontal scale is 
the time relative to perihelion (30.069 January 2015 UT) 
in days, minus the expected delay between peak outgassing and peak intensity 
due to beam dilution and opacity effect for a periodic outgassing
with similar periodicity. Production rates are apparent production rates 
computed in the stationary regime assumption. For \htwoo the larger error bar 
includes a 5\% uncertainty due to the 10\arcsec~ pointing uncertainty. 
\htwood uncertainty is dominated by statistical 
noise. The least square fit to the combined production rate (last line of 
Table~\ref{tabqper}) is overplotted.}
\label{figqper}
\end{figure}

\subsection{Periodic variation in \htwoo and \htwood production rates}
Since \htwoo and \htwood measurements with Odin were not simultaneous, but 
interlaced, in order to derive an accurate \htwoo/\htwood production rate 
ratio
it is necessary to determine if significant short-time variation of the 
activity is present. Inspection of production rates does suggest that time 
variation is present in the data. Since the observations of
\htwood constituted a series of ten successive Odin orbits, we did a time 
averaging of $\approx2.5$~h, corresponding to two consecutive orbits of
observations, yielding a sufficient signal-to-noise ratio (5--10) 
to look for time variations.
We fitted a sinusoidal time variation to the individual measurements 
of the production rates of \htwoo and \htwood, independently. We only 
selected the \htwoo measurements with pointing offsets less than 
$\sim$100\arcsec. The results are provided in Table~\ref{tabqper}. 
For both molecules, the sinusoidal fit is better than assuming no time variation
($\chi^2_\nu$ reduced by a factor $\sim$2.5), and the inferred period is 
similar ($T_p$=0.922-0.940 day). A simulation of periodic variations of the 
production rates anticipated a longer delay between peak outgassing and 
peak signal for \htwoo than \htwood (0.25 versus 0.14 day), which we took 
into account. We found different reference times $T_0$ for
\htwoo and \htwood, but these are strongly correlated to the $T_p$ value. 
Formally, the $T_p$-independent $1 \sigma$ uncertainty $\delta T_0$ is on 
the order of 0.4~days. We also tried  to fit the data 
with the same time dependence for \htwoo and \htwood,
using a fixed mean period of $T_p$ = 0.935 days (weighted average) and 
a corresponding reference time 31.742 Jan. 2015 UT. 
The goodness of the fits is similar 
(Table~\ref{tabqper}). A larger damping of the apparent 
production rate amplitude was expected for \htwoo due to optical depth 
effects. From the observations the effect is not so obvious: 
$\Delta~Q$/$Q_0$ is only slightly smaller for \htwoo than for \htwood 
(0.19 versus 0.24). We did
not investigate the possible variation of the gas temperature with the 
production rate, but this could explain a variation of the line intensity 
with time that was larger than anticipated.

We then combined both $Q_{\rm H_2^{16}O}$ and $Q_{\rm H_2^{18}O}$  
to derive our best estimate of the water production 
rate as a function of time during the period 30 January to 3 February 2015 
(last line of Table~\ref{tabqper}). 
Figure~\ref{figqper} shows the time evolution of 
the production rates measured for the two water isotopologues and best fit.

\subsection{Reference water production rates}

The reference water production rates were primarily derived from 
contemporaneous observations of the OH radical maser lines at 18cm 
with the Nan\c{c}ay radio telescope.
During the 12.8--16.8 January 2015 observing period of comet Lovejoy, 
both observations centered
on the comet and at 3.0\arcmin~ offset position were used to better constrain
the quenching of the maser inversion \citep{Ger98} yielding an average
water production rate $Q_{H_2O}=5.0\pm0.2\times10^{29}$molec.s$^{-1}$
\citep{Biv15}. For other periods and for comet Lemmon, the observations
of the OH radical could not be used to derive an accurate water production 
rate owing to significant uncertainties in the quenching of the maser 
inversion. We either used the water production rates reported elsewhere
\citep{Com14,Pag14} or made an interpolation from the Odin observations of
comet Lovejoy at perihelion.
Based on Nan\c{c}ay and Odin observations, 
we adopt a mean $Q_{\rm H_2O}=$ 5, 5.5, 
6, and $7\times10^{29}$~\mols~ for the 13--16 January, 13--25 January, 23--25 January,
and 26 January--3 February 2015 time intervals, respectively, with a $\pm20$\% time 
variation possible within each time interval. Observations done on a short 
time interval might be affected by a periodic time variation,
but this was tracked from the monitoring of the methanol production lines,
which are present in all IRAM observations settings (Table~\ref{tabsetup}).

\begin{table*}
\caption[]{Periodic variation of water production in comet Lovejoy}\label{tabqper}
\begin{center}
\begin{tabular}{lcccccc}
\hline
\multicolumn{7}{l}{Odin measurements 29 January -- 3 February 2015, $r_h=1.29$ AU} \\
\hline
Data points & Molecule & \multicolumn{4}{c}{Production rate variation fitted$^a$} & $\chi^2_\nu$ \\[0cm]
            &         &  $Q_0$ & $\Delta~Q$ & $T_p$   & $T_0$ & \\
            &         & \multicolumn{2}{c}{[$10^{27}$\mols]} & [days] & [UT day] \\
\hline
23$^b$ & \htwoo & $751.6\pm29.5$ & $140\pm21$ & $0.938\pm0.014$ & $31.797\pm0.038$-Jan. & 1.03\\
23$^b$ & \htwoo & $724.4\pm14.6$ & $141\pm22$ & $0.935^d$ & 31.742-Jan.$^d$ & 1.09\\
23$^b$ & \htwoo & $774.0\pm12.4$ & $0.0^c$ & -- & -- & 2.78\\
15     & \htwood & $1.524\pm0.065$ & $0.374\pm0.085$ & $0.922\pm0.031$ & $31.400\pm0.035$-Jan. & 1.21\\
15$^b$ & \htwood & $1.453\pm0.062$ & $0.314\pm0.083$ & $0.935^d$ & 31.742-Jan.$^d$ & 1.23\\
15     & \htwood & $1.421\pm0.055$ & $0.0^c$ & -- & -- & 2.44\\
38$^b$ & \htwoo, \htwood$^e$ 
                & $724\pm17$ & $147\pm19$ & $0.939\pm0.013$ & $31.912\pm0.022$-Jan. & 1.00\\
\hline
\end{tabular}
\end{center}
$^a$: $Q=Q_0+\Delta~Q\times\sin(2\pi(t-T_0)/T_p)$, where $T_p$ is the period in days
of the fitted sine variation and $T_0$ the reference time for the phase. \\
$^b$: including $5\%$ uncertainty due to a 10\arcsec~ pointing uncertainty. \\
$^c$: fit of constant production. \\
$^d$: fixed value (see text). \\
$^e$: Assuming \htwoo/\htwood = 500. \\
\end{table*}

\section{Isotopic ratios}
Table~\ref{tabareaqp} provides the list of molecular lines (main species
and isotopologues) observed at IRAM in comets C/2012~F6 (Lemmon) in 2013
and in C/2014~Q2 (Lovejoy) in 2015 and considered in this study.
Line intensities and production rates (or upper limits) derived using the 
models presented in Sect.~3 are given.
The corresponding isotopic ratios or $3~\sigma$ limits are presented in 
Table~\ref{tabisoratio}.

\begin{figure}
\centering
\resizebox{\hsize}{!}{\includegraphics[angle=0]{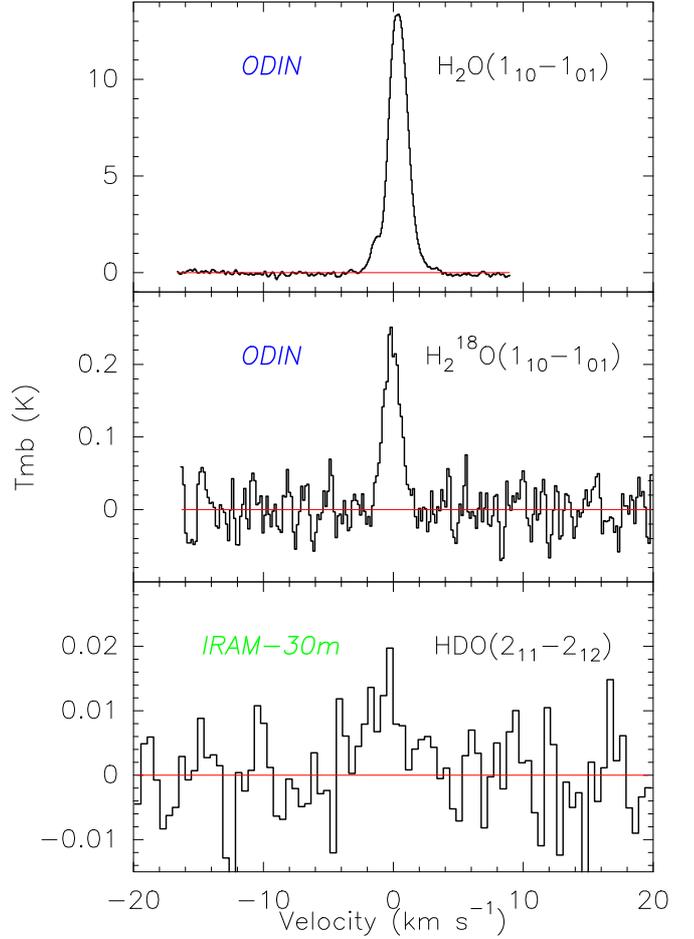}}
\caption{Comet C/2014~Q2 (Lovejoy):
Average on-nucleus spectra of the \htwoo~($1_{10}-1_{01}$) line at 556.936~GHz,
and \htwood~($1_{10}-1_{01}$) line at 547.676~GHz  obtained with Odin between
30 January and 3 February 2015. The spectrum of the HDO~($2_{11}-2_{12}$) line 
at 241.561~GHz observed with the IRAM 30m telescope on 13-24 January 2015 is shown below. 
While the \htwood and HDO lines are marginally blue-shifted 
($\Delta~v=-0.07$\kms~ in both cases) the \htwoo~($1_{10}-1_{01}$) line is asymmetric 
and red-shifted ($\Delta~v=+0.33$\kms~) due to self absorption by the cool gas
in the foreground of this optically thick line.
Vertical scale is main beam brightness temperature
and horizontal scale Doppler velocity in the comet rest frame.}
\label{figh2oisotopesq2}
\end{figure}

\begin{figure}
\centering
\resizebox{\hsize}{!}{\includegraphics[angle=270]{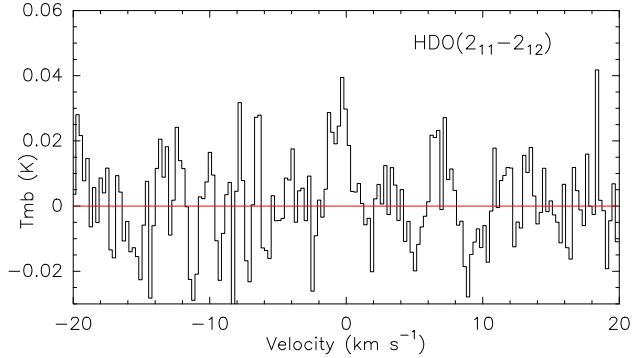}}
\caption{Comet C/2012 F6 (Lemmon): average on-nucleus spectrum of the 
HDO~($2_{11}-2_{12}$) line at 241.561~GHz observed with the IRAM 30m telescope
between 14 March and 8 April 2013. 
Vertical scale is main beam brightness temperature
and horizontal scale Doppler velocity in the comet rest frame.}
\label{figh2oisotopesf6}
\end{figure}

\subsection{\oratio ratio in comet Lovejoy}
The analysis of Odin observations in Sect.~4.2 (Table~\ref{tabqper}) 
enabled us to derive an accurate \htwoo/\htwood production rate ratio. 
From the sinusoidal fits and derived mean $Q_0$ production rates, we obtain:
\begin{itemize}
\item $Q_0$(\htwoo)/$Q_0$(\htwood) = $493\pm29$ for the independent fits;
\item $Q_0$(\htwoo)/$Q_0$(\htwood) = $499\pm24$ for the phased fits.
\end{itemize}
The second approach, which corresponds to a \htwoo/\htwood ratio that does not vary 
with time, yields  \oratio = $499\pm24,$ 
which is exactly the terrestrial value (498.7).

\subsection{D/H ratio}
Thanks to the wide frequency coverage of IRAM observations, lines of deuterated 
species of known cometary molecules were observed. None were detected
except the HDO~($2_{11}-2_{12}$) line at 241.562~GHz (Fig.~\ref{figh2oisotopesq2}). 
This is not surprising as water is over a hundred times more 
abundant than HCN, H$_2$S, or H$_2$CO. 
Nevertheless, the upper limits on the D/H ratio in H$_2$S or
H$_2$CO in comet Lovejoy are the lowest obtained so far in a comet.

\subsubsection{D/H ratio in water in comet C/2012 F6 (Lemmon)}
We observed the HDO~($2_{11}-2_{12}$) line at 
241.561~GHz in comet C/2012~F6 (Lemmon) in March-April 2013. 
The detection of HDO (Fig.~\ref{figh2oisotopesf6}) is also  marginal 
(4 $\sigma$) and the retrieved intensity was possibly
affected by poorer baselines on  8 April when we could
not use the telescope wobbler to cancel sky emission.
The inferred average HDO production rate for the period 14 March to 8 April 
2013 is $Q_{\rm HDO}=13\pm3\times10^{26}$\mols, yielding 
D/H = $6.5\pm1.6\times10^{-4}$. This is four times the Earth value
(VSMOW D/H=$1.56\times10^{-4}$), the highest value found in a comet, but
compatible with the Earth value at the $3~\sigma$ level.

\subsubsection{D/H ratio in comet C/2014~Q2 (Lovejoy)}
The tracking of the comet activity via monitoring of the CH$_3$OH production 
rate and the precise determination of the water outgassing rate using OH, 
\htwoo and \htwood  make us confident {\bf that we have}  a good 
H$_2$O production rate reference
in order to derive the D/H ratio. We report here the detection 
of HDO with the IRAM~30m telescope in this comet. The HDO~($2_{11}-2_{12}$) line was 
observed on 13, 15, 16, 23, and 24 January 2015. The average yields a 
$4~\sigma$ detection (Fig.~\ref{figh2oisotopesq2}) corresponding 
to $Q_{\rm HDO}=15.3\pm4.1\times10^{25}$\mols, 
hence a D/H ratio of $1.4\pm0.4\times10^{-4}$.
Although the detection is  marginal, in the worse case taking a 
$5~\sigma$ upper limit would yield $D/H < 1.86\times10^{-4}$, suggesting that 
comet Lovejoy has a D/H ratio for water similar to that of the Earth 
(VSMOW D/H=$1.56\times10^{-4}$), or possibly even lower.
Previous measurements of the D/H ratio in Oort cloud comets yielded values
larger than VSMOW (see Sect. 6.1).

\begin{figure}
\centering
\resizebox{\hsize}{!}{\includegraphics[angle=0]{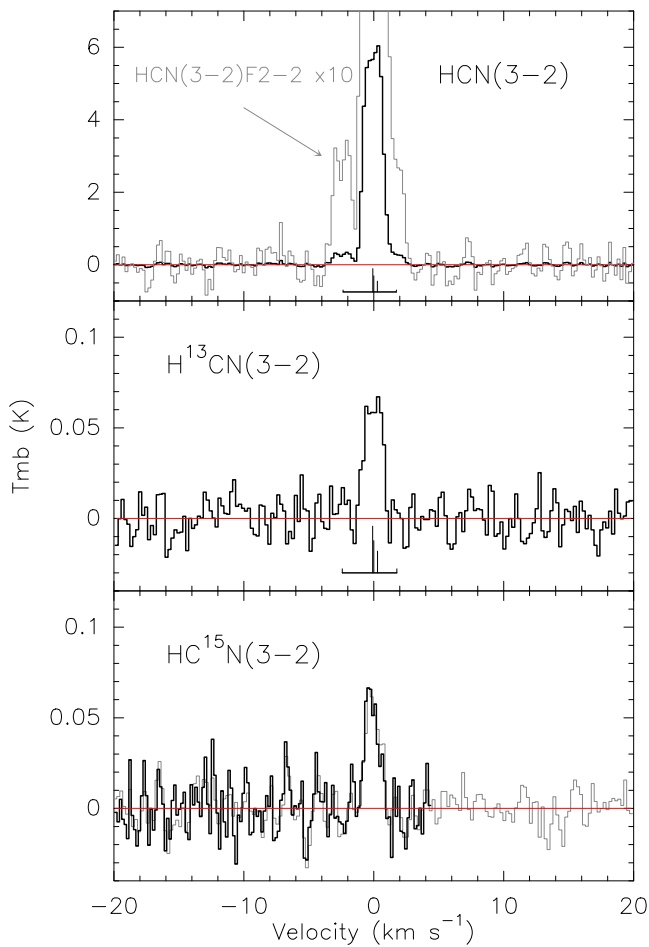}}
\caption{Comet C/2014~Q2 (Lovejoy):
Average on-nucleus spectra of the HCN~(3-2) line at 265.886~GHz, 
H$^{13}$CN~(3-2) line at 259.012~GHz, and HC$^{15}$N~(3-2) line at 258.157~GHz 
obtained with the IRAM 30m telescope on 13--24 January 2015. For HC$^{15}$N~(3-2) the
spectrum obtained with the VESPA autocorrelator (thick line) 
is superimposed on the FTS spectrum.
Vertical scale is main beam brightness temperature
and horizontal scale Doppler velocity in the comet rest frame.}
\label{fighcnisotopesq2}
\end{figure}

\begin{figure}
\centering
\resizebox{\hsize}{!}{\includegraphics[angle=0]{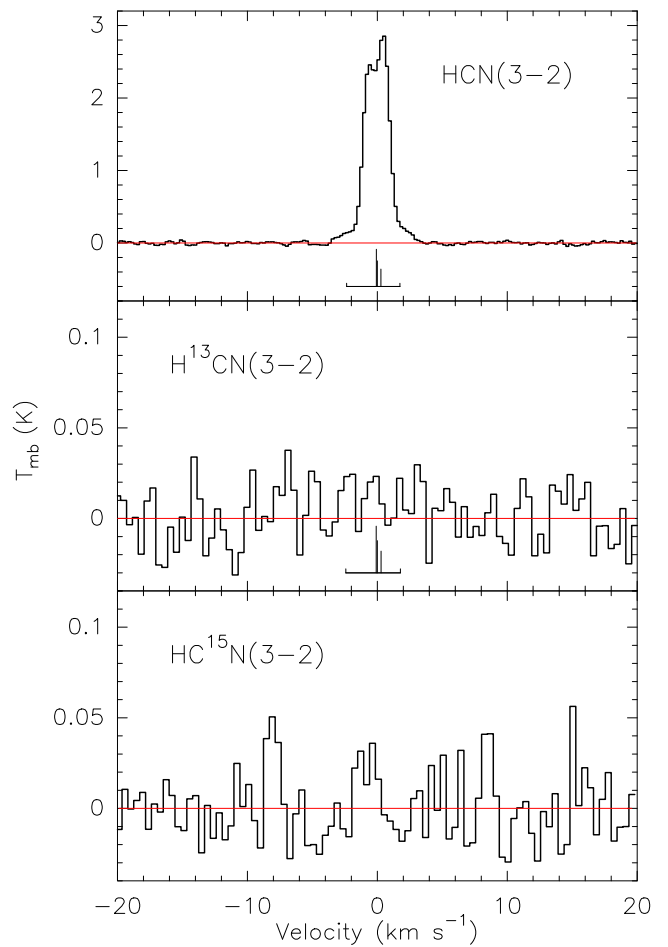}}
\caption{Comet C/2012 F6 (Lemmon): average on-nucleus spectrum of the HCN~(3-2) 
line at 265.886~GHz obtained with the IRAM 30m telescope between 14 March and 8 April 2013. 
Spectra covering the same velocity range obtained at similar times for 
H$^{13}$CN~(3-2) (259.012~GHz) and HC$^{15}$N~(3-2) (258.157~GHz) are also shown
for comparison, but do not show a clear detection.
Vertical scale is main beam brightness temperature
and horizontal scale Doppler velocity in the comet rest frame.}
\label{fighcnisotopesf6}
\end{figure}

\subsection{$^{14}$N/$^{15}$N ratio in HCN}
One of the primary goals of the observing program of comet
Lovejoy was to measure the $^{14}$N/$^{15}$N ratio in HCN.
On January 13, 15, and 16 we first observed the main isotopologue transition 
HCN~(3-2) at 265.886~GHz, then carried out longer integrations using the setup 
covering  the H$^{13}$CN~(3-2) line at 259.012~GHz and 
the HC$^{15}$N~(3-2) line at 258.157~GHz (Fig.~\ref{fighcnisotopesq2}).

The HCN reference production rate was computed taking  opacity
effects into account. The average value for the 13--16 January
period is $Q_{HCN}=4.87\pm0.02\times10\times10^{26}$~\mols, using  
pointing at the nucleus and at 5--8\arcsec offsets -- which both yield 
exactly the same value.
The HCN production rate based on the hyperfine
component HCN~($J=3-2,F=2-2$) with a total statistical weight of 3.7\% is 
marginally higher (+14\%, $2~\sigma$), possibly because of 
contamination by the wings of the main HCN hyperfine components.  

In order to minimize biases
due to time variation of the outgassing and other observing uncertainties 
(calibration, pointing uncertainty, etc.), we rescaled the HCN production rate to
the time when H$^{13}$CN~(3-2) and HC$^{15}$N~(3-2) were observed, using 
the CH$_3$OH production rate as a reference since several methanol lines 
were observed in each setup. 
The correction applied to $Q_{HCN}$ is +10\% for the 13--16 January 
observations and +27\% for 23--25 January. The compound  
HC$^{15}$N~(3-2) is clearly detected during both periods 
(signal-to-noise ratios of 9 and 6). The derived $^{14}$N/$^{15}$N ratios
are $145\pm16$ and $144\pm23$, for 13--16 January and 23--25 January, 
respectively. 

The compounds HC$^{15}$N~(3-2) and H$^{13}$CN~(3-2) were not clearly detected 
(Fig.~\ref{fighcnisotopesf6}) in comet C/2012~F6 (Lemmon), but the $2~\sigma$ 
value or lower limit we obtained for the $^{14}$N/$^{15}$N ratio in HCN 
($\geq 106$ or $152\pm72$) is fully compatible with values found in other
comets.

\subsection{$^{12}$C/$^{13}$C ratio}
The H$^{13}$CN~(3-2) line at 259.012~GHz is in the 
same spectral band as HC$^{15}$N~(3-2).
The production rates (assuming the same photo-dissociation lifetime as for HCN)
and relative abundances are derived in the same way for both molecules.
Owing to its hyperfine structure similar to that of H$^{12}$CN~(3-2) 
(Fig~\ref{fighcnisotopesq2}), the H$^{13}$CN~(3-2) line only contains 93\% of 
the flux in the (-1.2 -- 1.2~\kms) window and this was taken into account.

The derived $^{12}$C/$^{13}$C ratios in comet Lovejoy and Lemmon are
$109\pm14$ and $124\pm64$, respectively. Within the uncertainties,
they are compatible with the 
terrestrial value (89.7) although slightly higher ($1.4~\sigma$ for Lovejoy). 
This trend was also observed for comet 17P/Holmes \citep{Boc08} and
Hale-Bopp \citep{Jew97}. Combining the three \cratio measurements yields a 
value $2.5~\sigma$ larger than the Earth value. This interesting result 
needs to be further investigated in other comets. 

The sensitivity was not sufficient to detect  
other $^{13}$C isotopologues. The most sensitive
limit was reached for the {$^{13}$CH$_3$OH~($5_{+2}-4_{+1}E$) and  
$^{12}$CH$_3$OH~($5_{+2}-4_{+1}E$)} lines at 261.113 and 266.838~GHz,
respectively. The derived lower limit is
$^{12}$C/$^{13}$C $>61$ in comet Lovejoy, compatible with the Earth value.

\begin{figure}
\centering
\resizebox{\hsize}{!}{\includegraphics[angle=0]{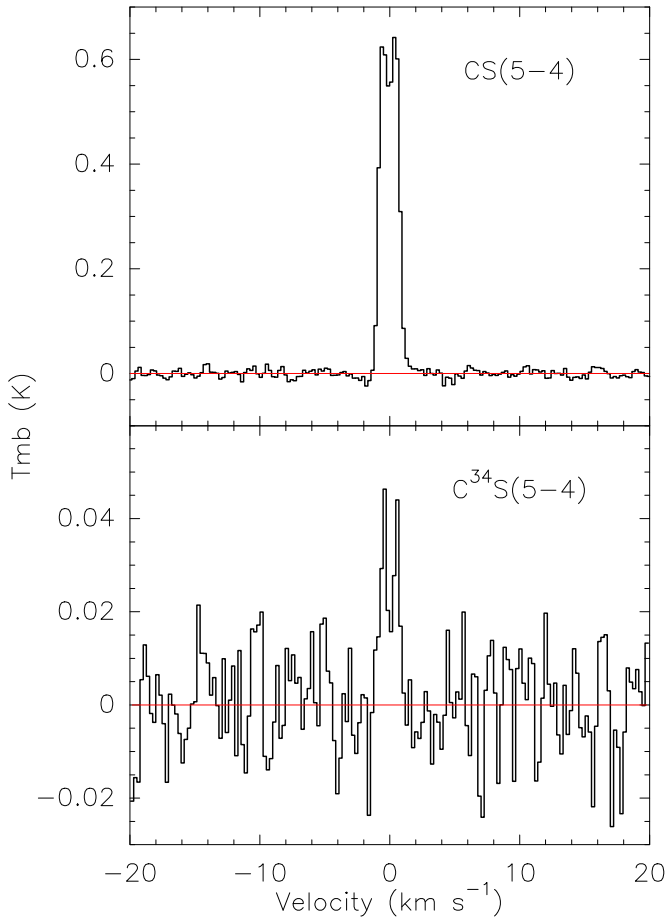}}
\caption{Comet C/2014~Q2 (Lovejoy):
Average on-nucleus spectra of the CS~(5-4) line at 244.936~GHz
and C$^{34}$S~(5-4) line at 241.016~GHz obtained with the IRAM 30m
telescope on 13-24 January 2015. Vertical scale is main beam brightness temperature
and horizontal scale Doppler velocity in the comet rest frame.}
\label{figcsisotopesq2}
\end{figure}

\begin{figure}
\centering
\resizebox{\hsize}{!}{\includegraphics[angle=0]{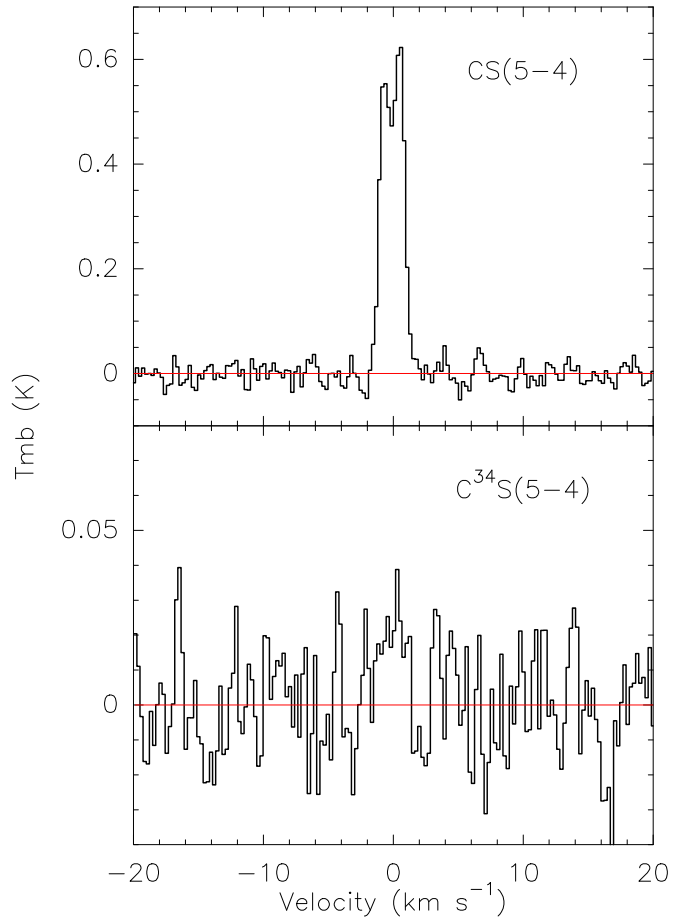}}
\caption{Comet C/2012 F6 (Lemmon): average on-nucleus spectra of the CS~(5-4) 
line at 244.936~GHz and C$^{34}$S~(5-4) line at 241.016~GHz obtained with the 
IRAM 30m telescope between 14 March and 8 April 2013. Vertical scale is main beam 
brightness temperature and horizontal scale Doppler velocity in the comet 
rest frame.}
\label{figcsisotopesf6}
\end{figure}

\subsection{$^{32}$S/$^{34}$S ratio}
The wide 8~GHz frequency coverage of the EMIR receiver at IRAM, 
allows us to observe both C$^{32}$S and C$^{34}$S 
or H$_2^{32}$S and H$_2^{34}$S lines in the same FTS spectrum 
since they are less than 4~GHz apart.
As a consequence, the isotopic ratios do not depend on any calibration 
or pointing effects and, since the lines are mostly optically thin, 
the abundance ratio is very close to the line intensity ratio.
The integration time on the H$_2$S lines (either around 168 or 216 GHz) was 
not sufficient to yield a significant constraint on the  H$_2^{32}$S/H$_2^{34}$S
ratio. On the other hand, the C$^{34}$S~(5-4) line at 241016.089~MHz was detected
in both comets (Figs.~\ref{figcsisotopesq2} and \ref{figcsisotopesf6}). 
The accuracy of the $^{32}$S/$^{34}$S ratio in CS is given by the 
signal-to-noise ratio of the C$^{34}$S line, and we found 
$^{32}$S/$^{34}$S = $20\pm5$ and $24.7\pm3.5$ in comets Lemmon and 
Lovejoy, respectively. We do not
find any significant departure from the Earth value of $^{32}$S/$^{34}$S=22.5.
The C$^{33}$S line lies  between the C$^{34}$S and C$^{32}$S lines. The upper 
limit for the C$^{33}$S~(5-4) line at 242.913~GHz in comet Lovejoy 
yields $^{32}$S/$^{33}$S$>50$. This is the first reported estimate of
$^{33}$S in a comet, but this is only a lower limit compatible with the 
Earth value (126.7).

\begin{table*}
\caption[]{Line intensities from IRAM observations and production rates}\label{tabareaqp}
\begin{center}
\begin{tabular}{lllccc}
\hline
Date & Molecule & Transition & Frequency & Intensity & Total production rate \\
$[yyyy/mm/dd.d-dd.d]$ &  &   & [MHz]  & [mK~\kms] & [$10^{26}$~molec.s$^{-1}$] \\
\hline
\multicolumn{5}{l}{Comet C/2012~F6 (Lemmon) March-April 2013} \\
\hline
2013/03/14.5-18.6 & HCN & $3-2$ & 265886.432 & $6055\pm36$ & $15.3\pm2.1^a$ \\
2013/04/06.46     & HCN & $3-2$ &            & $5966\pm50$ & $13.1\pm0.8^a$ \\
2013/03/14.5-18.6 & H$^{13}$CN & $3-2$ & 259011.798 & $65\pm44$ & $<0.28$ \\
2013/04/06.56     & H$^{13}$CN & $3-2$ &            & $43\pm32$ & $<0.20$ \\
2013/03/14.5-37.6 & H$^{13}$CN & $3-2$ &            & $51\pm26$ & $0.12\pm0.06$ \\
2013/03/14.5-18.6 & HC$^{15}$N & $3-2$ & 258156.996 & $108\pm36$ & $<0.25$ \\
2013/04/06.46     & HC$^{15}$N & $3-2$ &            &  $11\pm22$ & $<0.14$ \\
2013/03/14.5-37.6 & HC$^{15}$N & $3-2$ &            &  $40\pm19$ & $0.09\pm0.04$ \\
2013/03/15.6-38.6 & DCN & $3-2$ & 217238.538 &  $<133$ & $<0.52$ \\

2013/04/08.47  & H$_2$S & $1_{10}-1_{01}$ & 168762.762 & $234\pm36$ & $77\pm13^a$ \\
2013/04/08.47  & H$_2^{34}$S & $1_{10}-1_{01}$ & 167910.516 & $<66$ & $<21$ \\

2013/03/14.6-18.6 & CS  & $5-4$ & 244935.557 & $1477\pm38$ & $14.7\pm0.6^a$ \\
2013/04/06.56     & CS  & $5-4$ &            & $1019\pm25$ &  $9.5\pm0.2$ \\
2013/03/14.6-18.6 & C$^{34}$S & $5-4$ & 241016.089 & $50\pm31$ & $0.51\pm0.31$ \\
2013/04/06.56     & C$^{34}$S & $5-4$ &            & $58\pm16$ & $0.57\pm0.16$ \\
2013/03/15.6-38.6 & C$^{34}$S & $5-4$ &            & $53\pm13$ & $0.54\pm0.13$ \\

2013/03/14.6-18.6 & HDO & $2_{11}-2_{12}$ & 241561.550 & $88\pm29$ & $21\pm7$ \\
2013/04/06.56     & HDO & $2_{11}-2_{12}$ &            & $44\pm19$ & $12\pm4$ \\
2013/03/14.6-37.6 & HDO & $2_{11}-2_{12}$ &            & $54\pm14$ & $13\pm3$ \\
\hline
\multicolumn{6}{l}{Comet C/2014~Q2 (Lovejoy) 13--25 January 2015} \\
\hline
2015/01/13.7-16.8 & HCN & $3-2$ & 265886.432 & $10848\pm39$ & $4.9\pm0.2^a$ \\
2015/01/13.7-16.8 & H$^{13}$CN & $3-2$ & 259011.798 & $114\pm13$ & $0.049\pm0.006$ \\
2015/01/23.73     & H$^{13}$CN & $3-2$ &            &  $86\pm18$ & $0.043\pm0.009$ \\
2015/01/13.7-23.7 & H$^{13}$CN & $3-2$ &            & $100\pm12$ & $0.046\pm0.006$ \\
2015/01/13.7-16.8 & HC$^{15}$N & $3-2$ & 258156.996 &  $86\pm 9$ & $0.037\pm0.004$ \\
2015/01/23.73     & HC$^{15}$N & $3-2$ &            &  $85\pm13$ & $0.043\pm0.007$ \\
2015/01/13.7-23.7 & HC$^{15}$N & $3-2$ &            &  $83\pm 7$ & $0.039\pm0.003$ \\
2015/01/13.7-24.7 & DCN & $3-2$ & 217238.538 &  $<39$ & $<0.031$ \\
2015/01/13.7-16.8 & CH$_3$OH & $5_2-4_1$E & 266838.123 & $1345\pm24$ & $117\pm3^a$ \\
2015/01/13.7-23.7 & $^{13}$CH$_3$OH & $5_2-4_1$E & 263113.343 & $<23$ & $<2.4$ \\
2015/01/13.7-25.7 & H$_2$CO & $3_{13}-2_{12}$ & 211211.469 & $978\pm11$ & $180\pm10^a$ \\
2015/01/13.7-25.7 & H$_2$CO & $3_{12}-2_{11}$ & 225697.772 & $1181\pm17$ & $180\pm10^a$ \\
2015/01/13.7-23.7 & HDCO & $4_{14}-3_{13}$ & 246924.600 & $<18$ & $<0.38$ \\
2015/01/13.7-23.7 & HDCO & $4_{04}-3_{03}$ & 256585.531 & $<21$ & $<0.36$ \\

2015/01/13.8-25.7 & H$_2$S & $2_{20}-2_{11}$ & 216710.437 & $227\pm12$ & $17.3\pm0.9^a$ \\
2015/01/13.8-25.7 & H$_2^{34}$S & $2_{20}-2_{11}$ & 214376.924 & $16\pm11$ & $<2.6$ \\
2015/01/13.8-23.7 & HDS & $2_{11}-2_{02}$ & 257781.410 & $<21$ & $<0.6$ \\
2015/01/13.8-16.9 & CS  & $5-4$ & 244935.557 & $1049\pm10$ & $2.14\pm0.02^a$ \\
2015/01/23.7-24.7 & CS  & $5-4$ &            & $1089\pm10$ & $2.51\pm0.02^a$ \\
2015/01/13.8-23.7 & CS  & $5-4$ &            & $1076\pm 7$ & $2.30\pm0.02^a$ \\
2015/01/13.8-16.9 & C$^{34}$S & $5-4$ & 241016.089 & $37\pm9$ & $0.077\pm0.019$ \\
2015/01/23.73     & C$^{34}$S & $5-4$ &     & $53\pm13$ & $0.126\pm0.031$ \\
2015/01/13.8-23.7 & C$^{34}$S & $5-4$ &     & $42\pm 6$ & $0.093\pm0.013$ \\
2015/01/13.8-23.7 & C$^{33}$S & $5-4$ & 242913.610 & $<21$     & $<0.046$ \\

2015/01/13.8-16.9 & HDO & $2_{11}-2_{12}$ & 241561.550 & $31\pm 8$ & $1.8\pm0.4$ \\
2015/01/23.7-24.7 & HDO & $2_{11}-2_{12}$ &            & $23\pm11$ & $1.6\pm0.7$ \\
2015/01/13.8-24.7 & HDO & $2_{11}-2_{12}$ &            & $25\pm 7$ & $1.5\pm0.4$ \\
\hline
\end{tabular}
\end{center}
\noindent{$^a$: Average production rate for the period also taking into
account  measurements at 5 to 20\arcsec~ offsets. For formaldehyde
the offset positions were used to constrain the contribution of the
distributed source: 80\% with a scale-length of 10\,000 km.}\\
\end{table*}

\begin{table}
\caption[]{Isotopic ratios in comet C/2014~Q2 and C/2012~F6}\label{tabisoratio}
\begin{center}
\begin{tabular}{ccc}
\hline
Isotopic ratio & Molecule & value \\
\hline
\multicolumn{3}{c}{Comet C/2012~F6 (Lemmon) Mar.-Apr. 2013} \\
\hline
D/H               & H$_2$O & $6.5\pm1.6\times10^{-4}$  \\
                  & HCN    & $<0.045$                 \\
$^{12}$C/$^{13}$C & HCN    & $124\pm64$ or $\geq89^b$  \\
$^{14}$N/$^{15}$N & HCN    & $152\pm72$ or $\geq106^b$   \\
$^{32}$S/$^{34}$S & CS     & $20\pm5$                \\
                  & H$_2$S & $>3.5$                 \\
\hline
\multicolumn{3}{c}{Comet C/2014~Q2 (Lovejoy) Jan. 2015} \\
\hline
D/H               & H$_2$O & $1.4\pm0.4\times10^{-4}$ \\
                  & HCN    & $<0.006$                 \\
                  & H$_2$S & $<0.017$                \\
                  & H$_2$CO & $<0.007$                \\
$^{12}$C/$^{13}$C & HCN    & $109\pm14$              \\
                  & CH$_3$OH & $>61$                 \\
$^{14}$N/$^{15}$N & HCN    & $145\pm12$              \\
$^{16}$O/$^{18}$O & H$_2$O & $499\pm24^a$             \\
$^{32}$S/$^{34}$S & CS     & $24.7\pm3.5$            \\
                  & H$_2$S & $>7$                    \\
$^{32}$S/$^{33}$S & CS     & $>50$                   \\
\hline
\end{tabular}
\end{center}
Note: $^a$: From time variable fits of the production rates.\\
      $^b$: based only on the backends with the lowest noise.\\
\end{table}

\begin{figure}
\centering
\resizebox{\hsize}{!}{\includegraphics[angle=270]{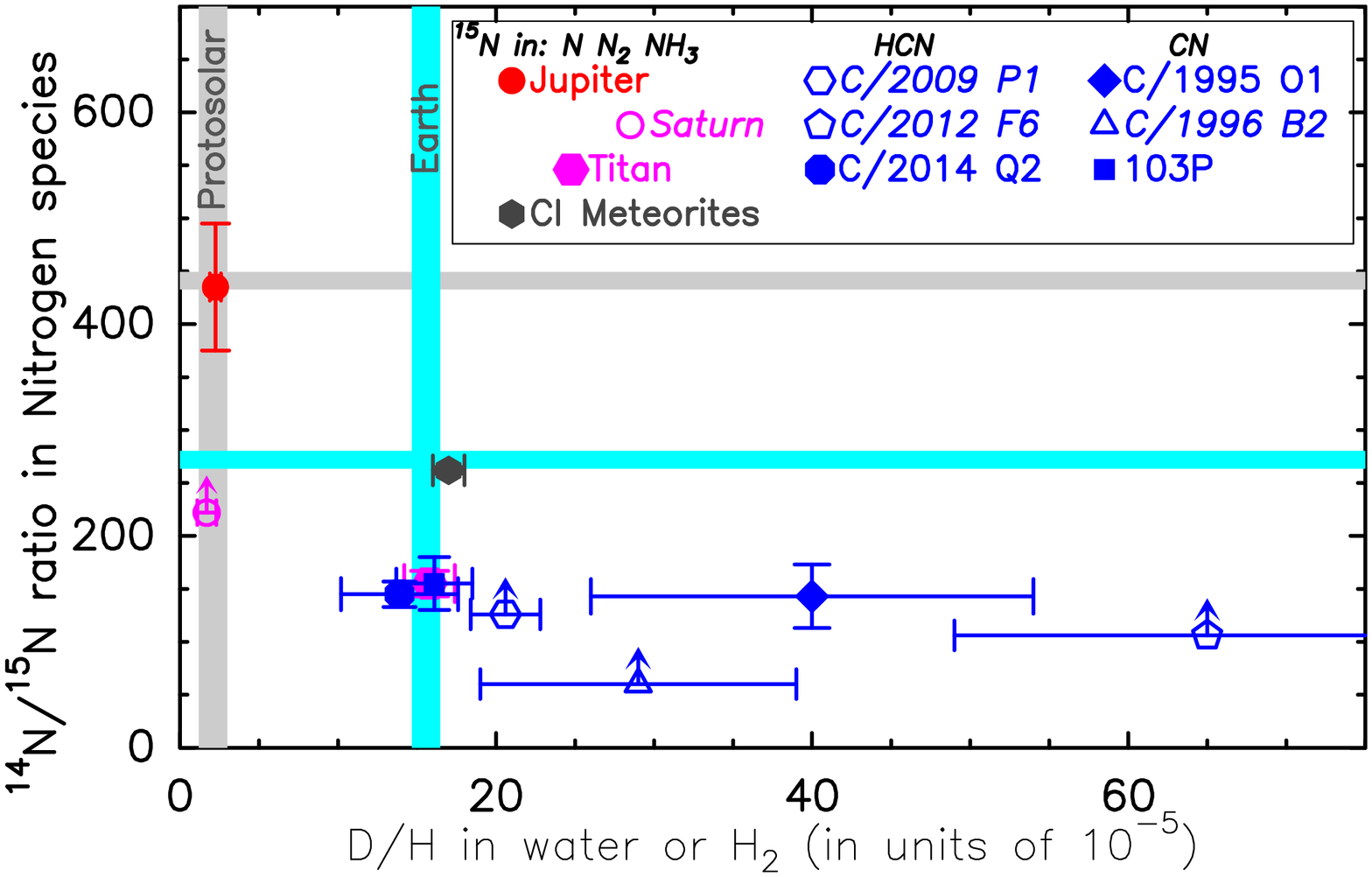}}
\caption{D/H ratios and \nratio ratios in solar system objects,
compared to the protosolar (gray bands) and Earth (blue bands) values 
\citep{Rob00, Mar11}. Jupiter and Saturn values are from 
\citet{Mar11}, \citet{Fur15}, and \citet{Fle14}; Titan values from 
\citet{Abb10}, \citet{Nie10}, and \citet{Man09}; meteorites from 
\citet{Wai09} and \citet{Ker85}; and comet values from 
\citet{Boc98}, \citet{Mei98}, \citet{Cro04a}, \citet{Har11}, \citet{Boc12}, 
\citet{Boc08}, \citet{Hut09}, and this paper. 
The symbols corresponding to each source are given in the enclosed 
legend box. They are  ordered according to the species in which the 
\nratio~ was measured from left to right: atomic nitrogen, N$_2$ 
(as for Earth), ammonia, HCN, and CN.
Empty symbols with names in italics are for lower limits on the \nratio.
}
\label{figdh15n}
\end{figure}

\section{Discussion}
We measured a terrestrial value for the \oratio ratio 
in water in comet Lovejoy, as was the case in all comets in which this 
ratio had been measured before \citep{Bal95,Biv07,Boc12,Alt15,Boc15}.

The D/H and \nratio ratios in solar system bodies are of particular importance 
as they vary from object to object, with the heavier elements enriched in the
Earth and small bodies compared to Jupiter or the protosolar values.
These ratios trace the various enrichment and fractionation mechanisms that
were at play in the proto-planetary nebula.
The other isotopic ratios investigated here do not show significant
variations in solar system objects within the accuracy of our
measurements. Figure~\ref{figdh15n} shows the \nratio versus D/H ratios
measured in solar system objects, including our new measurements
presented here.   

\subsection{{\bf Earth-like water in comet Lovejoy}}
We obtained a measurement or at least a stringent upper limit for the
D/H ratio in water of two dynamically old Oort cloud comets.
We measured the lowest D/H ratio in water in a comet so far: 
$1.4\pm0.4\times10^{-4}$ ($< 1.8\times10^{-4}$; {\bf $5~\sigma$}) in Lovejoy, 
while the corresponding ratio could be four times higher in
comet Lemmon.
These two D/H measurements are based on the same line in two comets of 
similar activity. Comet Lemmon was slightly more active but also two times  
farther away. We do not expect that the excitation mechanism of HDO in the 
coma of these comets can yield a factor of four difference in the derived 
production rates. So the observed diversity in the D/H ratio
can be real and would suggest that these two comets, although 
they come from the same current reservoir, were formed in different places 
or at different times in the young solar system. 
In either case, this brings additional proof that comet dynamical origin does 
not imply a specific formation region in the disk.
Previous studies 
\citep{Har11,Lis13,Boc12,Alt15} have shown a growing diversity in the
measured values of the D/H ratio in cometary water, and especially 
variations by more than a factor of 3 among Jupiter family comets,
which are thought to originate from the Kuiper belt.
Now we find that a similar range of values
is present among Oort cloud comets.
This is in line with the latest models \citep{Bra13}, which suggest that all 
comets were formed in the same extended massive original Kuiper Belt beyond 
the giant planets and were later scattered to the two current reservoirs, 
the scattered Kuiper disk and the Oort cloud. The diversity in deuterium enrichment
would occur because  the pre-solar cloud was initially enriched in 
HDO (higher D/H in water than in H$_2$: protosolar D/H value in H$_2$ is 
$2\times10^{-5}$, Fig.\ref{figdh15n}), but the inner warmer parts of the
proto-planetary nebula lost more of their initial enrichment in 
deuterated water.   

\subsection{\nratio in comets}
Measurements of  \nratio in comets are relatively scarce. Although the
\nratio\ ratio is 5--20 times lower than the \htwoo/HDO ratio, the nitrogen bearing
molecules in cometary comae are less than 1\% in abundance relative to
water. So the detection of $^{15}$N isotopologues requires very active comets.
The most sensitive technique is optical high resolution spectroscopy,
which has provided several detections of C$^{15}$N \citep{Manf05,Manf09,Hut09}
and recently detections of $^{15}$NH$_2$ \citep{Rou14,Shi15}, but this
technique
only gives access to daughter species. The compounds N$_2$ and NH$_3$ are 
not very abundant \citep{Rub15,Biv12} and  do not have strong lines observable
from the ground, nor do their $^{15}$N isotopologues. Only HCN is easily
detectable in the radio and infrared. We report here the third 
clear (S/N > 10) detection of HC$^{15}$N in a comet after its detection in
comets Hale-Bopp and 17P/Holmes \citep{Boc08}.

We measure the same enrichment in $^{15}$N in HCN in comet Lovejoy 
as in other comets, in spite of their different dynamical origins 
\citep{Boc08}. The same enrichment was also found in CN in many other comets 
\citep{Hut09}. This is a factor of two enrichment in $^{15}$N in comparison 
to the Earth atmospheric value (272) and a factor of three in comparison 
to the proto-planetary value \citep[441,][]{Mar11} (Fig.~\ref{figdh15n}). 
The enrichment is similar to that found in NH$_2$ \citep{Rou14,Shi15}, 
thought to be mostly produced by the photo-dissociation of NH$_3$, 
the main volatile carrier of cometary nitrogen \citep{Biv12}.

Correspondingly low \nratio ratios are also detected in carbonaceous IDPs
\citep{Mes00,Flo04}. These large enrichments with respect to solar or
terrestrial values can be explained by interstellar chemistry theories 
involving ion-molecule $^{15}$N fractionation at 10~K \citep[e.g.,][]{Wir12}.
Recently, significant nitrogen fractionation has been measured in dark cloud 
cores with ratios as low as $150$ for nitriles \citep{Mil15}. Thus, 
the similarity of the \nratio ratios found in comets and IDPs strengthens 
a putative link to interstellar chemistry as the origin of isotopically 
anomalous organic particles in comets.

Nevertheless, our observations confirms that the trend of a twofold enrichment 
in $^{15}$N compared to Earth observed in C$^{15}$N is also present in HCN.
A consequence concerning the debate of the origin of CN in cometary coma is 
that we cannot exclude HCN as the sole parent of CN on the basis of
different \nratio ratios.

\section{Conclusions}
Chemical diversity is observed in the population of Oort Cloud
comets, with abundances varying by up to a factor of ten for
several species. The origin of this diversity is unclear and 
might reflect comet formation at different places {\bf or} 
times in the early solar system. 

The observations presented here
provide further evidence that diversity is present in the
enrichment in deuterium in cometary water with respect to the 
protosolar value (HD/H$_2$), both in Jupiter family and 
Oort cloud comets, in line with the 
latest scenarios of comet origins \citep{Bra13}. 
On the one hand, we found Earth-like D/H, \oratio, \cratio, and 
\sratio ratios in comet C/2014~Q2 (Lovejoy), strengthening the role 
that some comets may have played in supplying material to the young
Earth, especially complex organic molecules \citep{Biv15}.
On the other hand we confirm the trend of finding a systematic
twofold enrichment in $^{15}$N in cometary HCN relative to its abundance
in Earth N$_2$, whose origin is puzzling.

We were also able to obtain in comet C/2014~Q2 sensitive limits, some of the
best obtained so far with remote observations, on isotopic ratios in other molecules 
such as D/H in H$_2$S, H$_2$CO, or C$^{32}$S/C$^{33}$S.

\begin{acknowledgements}
The IRAM observations were conducted under the target of opportunity 
proposal D04-14 and regular proposal 128-14 and we gratefully 
acknowledge the support from the IRAM director for awarding us
discretionary time and the IRAM staff for its support and for scheduling
the observations on short notice.
This research has been supported by the Programme national de 
plan\'etologie de l'Institut des sciences de l'univers (INSU).
SNM acknowledges the NASA Planetary Astronomy program for support.

\end{acknowledgements}


\end{document}